\documentstyle[epsfig]{article}
\begin{document}
\large 
\begin{center}
{\Large \bf {Two-Stream Instability as a Mechanism for Toroidal
Magnetic Field Generation in the Magnetosphere of Crab Pulsar}}
\end{center}
\vskip 15pt
\begin{center}
{\bf Irakli S. Nanobashvili}
\end{center}
\vskip 15pt \noindent
{\it Institute
of Theoretical Physics, Ilia State University, Tbilisi, Georgia \\
Andronikashvili Institute of Physics, Iv. Javakhishvili Tbilisi
State University, Tbilisi, Georgia} \vskip 15pt \noindent {E-mail
address: inanob@yahoo.com} \vskip 15pt \noindent {\bf Abstract.}
New plasma mechanism for the generation of toroidal magnetic field
in the magnetosphere of Crab pulsar is presented. It is based on
the development of two-stream instability in the relativistic
electron-positron plasma of the pulsar magnetosphere. In
particular, pulsar magnetosphere relativistic plasma is penetrated
by ultrarelativistic electron beam and two-stream instability
develops, as a result of which toroidal magnetic field is
gene\-rated.

\vskip 15pt \noindent {\bf 1. Introduction} \vskip 15pt Nowadays
it is widely accepted that pulsars are rapidly rotating neutron
stars [1-4] with strong magnetic field of the order
$10^{12}{\div}10^{13}$ G. Neutron star is surrounded with
magnetosphere, which is filled with re\-lativistic
electron-positron plasma (see example [5]). Pulsar radiation is
generated in its magnetosphere most probably as a result of the
development of different plasma processes in the region above the
pulsar magnetic poles ([6]). In order to find the processes which
are responsible for the generation of pulsar radiation it is
essential to know in detail the structure of the magnetosphere
where these processes develop.

In the pulsar magnetosphere, close to its surface, magnetic field
has a dominant role - its energy exceeds the energy of the
magnetospheric relativistic plasma by many orders of the
magnitude. Magnetic field of pulsar has the dipole structure. It
is frozen in magnetospheric plasma and in pulsar too. Therefore,
solid body type rotation - corotation of pulsar, its magnetic
field and magnetospheric relativistic plasma takes place. In the
region of the magnetosphere where the magnetic field lines are
closed magnetospheric plasma is confined by the magnetic field and
it can not leave the magnetosphere. In this region we have "quiet
corotation" if one can say so. Plasma can leave the magnetosphere
only from the conical region (with small angle of opening) above
the pulsar magnetic poles. In this region magnetic field lines are
"opened" and since plasma particles follow these lines they leave
pulsar magnetosphere and form relativistic pulsar wind. In case of
Crab pulsar opened magnetic field lines practically lie in the
equatorial plane of rotation because pulsar magnetic axis is
nearly perpendicular to its rotation axis [7,8]. In general,
opened magnetic field lines of pulsar are considered as almost
straight radial lines in the region close to its surface, because
in this region their curvature is small. Besides, in this region
we have rigid corotation - plasma particles rotate together with
the magnetic field lines and also move along them. It is evident
that this can not take place on large radial distance. In
particular, corotation is strictly impossible beyond the light
cylinder (cylindrical surface on which the corotation velocity
equals to the speed of light). At the same time in this region,
which is called wind zone, we have not the magnetospheric
relativistic plasma but the relativistic pulsar wind. In the wind
zone magnetic field is practically purely toroidal, it is still
frozen in plasma, but its energy is smaller then the energy of the
relativistic pulsar wind. From all the above mentioned it follows
that somewhere in the pulsar magnetosphere - inside the light
cylinder - toroidal magnetic field must be generated and
corotation must be violated.

In the present paper one possible plasma mechanism for the
gene\-ration of toroidal magnetic field in the magnetosphere of
Crab pulsar is suggested. In the forthcoming section pulsar
magnetosphere structure before the gene\-ration of toroidal
magnetic field is discussed. In third section the mechanism of
toroidal magnetic field generation in the pulsar magnetosphere is
presented. The mechanism is based on the development of two-stream
instability in the magnetospheric relativistic electron-positron
plasma.

\vskip 15pt \noindent {\bf 2. The Structure of the pulsar
magnetosphere before the generation of toroidal magnetic field}
\vskip 15pt As it has been already mentioned above pulsar
magnetosphere is filled with relativistic electron-positron
plasma. This plasma appears there as a result of cascade process
which develops in the following way. Since matter inside pulsar is
in superconductive state magnetic field is frozen in pulsar and
rotates together with it. As a result of this rotation electric
field is generated which extracts charged particles from pulsar
surface [9]. Depending of the direction of generated electric
field the particles extracted from the pulsar surface may be
electrons [10], or positrons [11] and ions [12]. Here it will be
assumed that charged particles extracted from the pulsar surface
by electric field are electrons. In the pulsar magnetosphere
electrons are accelerated by electric field and acquire
ultrarelativistic velocities. Electrons follow the magnetic field
lines and have only the longitudinal (with res\-pect to the
magnetic field line) component of velocity, because perpendicular
component is lost in the strong magnetic field of pulsar after the
rapid radiation with synchrotron mechanism. Since magnetic field
lines are curved, the electrons moving along them with
ultrarelativistic velocities radiate curvature radiation
$\gamma$-quanta. Then, in the strong magnetic field of pulsar
$\gamma$-quanta decay into electron-positron pairs. These
electrons and positrons are also accelerated by electric field and
emit  curvature radia\-tion $\gamma$-quanta, which again decay
into electron-positron pairs etc. This cascade process leads to
the formation of dense relativistic electron-positron plasma in
the pulsar magnetosphere [5] . This plasma is penetrated by
primary ultrarelativistic electron beam.

\begin{figure}{}
\begin{center}
\includegraphics[width=7cm]{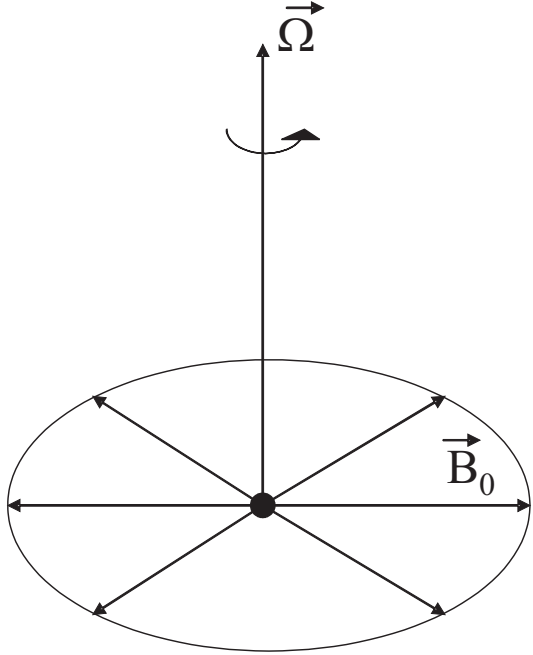}
\end{center}
\caption{Geometric model of crab pulsar magnetic field.}
\end{figure}

Now about the simplified geometric model of Crab pulsar magnetic
field which will be used below. As it has been already mentioned
above rotation axis and magnetic axis of Crab pulsar are nearly
perpendicular. So, pulsar magnetic field lines are considered as
radial straight lines located in the equatorial plane of rotation
(see the Fig. 1). This assumption is justified and at the same
time this is not an approximation of monopolar magnetic field for
the following reasons: first of all only the open magnetic field
lines which come out from one magnetic pole of pulsar are
discussed (all the results obtained below will be the same for the
open magnetic field lines which come out from ano\-ther magnetic
pole of pulsar, just the direction of these magnetic field lines
will be the opposite). Besides, these magnetic field lines are
discussed in the thin layer of the magnetosphere close to pulsar
surface. The thickness of this layer (${\approx}10^7$ cm) is much
less than the curvature radius of Crab pulsar magnetic field lines
(${\approx}10^8$ cm), therefore in this layer magnetic field lines
can be considered straight. The reason why one gets for the
magnetic field the picture seen on Fig. 1 is the rotation of those
group of magnetic field lines which have been just defined above.

\vskip 15pt \noindent {\bf 3. Two-stream instability development
and toroidal magnetic field gene\-ration in the pulsar
magnetosphere} \vskip 15pt As we have seen above relativistic
electron-positron plasma in the pulsar magnetosphere is penetrated
by ultrarelativistic electron beam. As a result of the
inte\-raction of ultrarelativistic electron beam with relativistic
plasma of the pulsar magnetosphere two-stream instability may
develop. The possibility of the generation of pulsar radiation as
a result of two-stream instability development has been studied in
the papers [11-19]. In the present paper the possibility of
toroidal magnetic field generation in the magnetosphere of Crab
pulsar as a result of two-stream instability deve\-lopment is
investigated. For this purpose the standard set of equations
describing the dynamics of cold relativistic magnetized plasma is
used:

$$
{{{\partial}{\vec p}_{(\alpha)}}\over {{\partial}t}}+\left({\vec
V}_{(\alpha)}{\vec {\nabla}}\right){\vec
p}_{(\alpha)}={{e_{(\alpha)}}\over {m}}\left({\vec E}+\left[{\vec
V}_{(\alpha)}{\times} {\vec B}\right]\right), \eqno (1)
$$

$$
{{{\partial}{n_{(\alpha)}}}\over {{\partial}t}}+div
\left(n_{(\alpha)}{\vec V}_{(\alpha)}\right)=0, \eqno (2)
$$

$$
rot{\vec E}=-{{{\partial}{\vec B}}\over {{\partial}t}}, \eqno (3)
$$

$$
rot{\vec B}=4{\pi}{\vec j}+{{{\partial} {\vec E}}\over
{{\partial}t}}, \eqno (4)
$$
where $e_{(\alpha)}$ and $m$ are particle electric charge and mass
respectively, ${\vec V}_{(\alpha)}$ and ${\vec
p}_{(\alpha)}={{\gamma}_{\alpha}}{m}{\vec V}_{(\alpha)}$
(${{\gamma}_{\alpha}}$ being particle Lorentz-factor) are particle
tree-velocity and momentum, $n_{(\alpha)}$ is the particle
density, $\vec j$ is the current density and $\vec E$ and $\vec B$
are electric and magnetic fields. Subscript $(\alpha)$ denotes the
group of particles (we have two groups - plasma ($\alpha=1$) and
beam ($\alpha=2$)). In the equations (1-4) so-called "geometric"
unites - $c=G=1$ are used and momentum is changed by normalized
momentum ${\vec p}{\rightarrow}{{\vec p}/{m}}$.

The dynamics of electromagnetic perturbations in the system
plasma-beam is studied. These studies are performed in the
reference frame of a rotating magnetic field line (the geometry of
the magnetic field lines being discussed in the previous section).
At the same time the re\-ference frame in which the investigations
are performed is moving radially outwards along the magnetic field
line with such a constant velocity that in this frame the
velo\-cities of plasma and beam are equal ($\vert{\vec
V}_{(1)}\vert=\vert{\vec V}_{(2)}\vert=V$) and directed in
opposite direction (beam velocity ${\vec V}_{(2)}$ being directed
radially outwards and plasma velocity ${\vec V}_{(1)}$ - radially
inwards). Here we discuss the dynamics of the perturbations wave
vector of which is directed along the x-axis (x-axis is parallel
to the pulsar rotation axis) - ${\vec k}\left(k_x,0,0\right)$,
perturbed electric field is directed opposite to the z-axis
(z-axis being directed radially outwards from pulsar) - ${\vec
{E_1}}\left({0,0,-E_{1z}}\right)$, unperturbed magnetic field
$B_0$ (which is pulsar magnetic field) is also directed along the
z-axis and perturbed magnetic field has only the toroidal -
y-component - ${\vec {B_1}}\left(0, B_{1y},0\right)$, see the Fig.
2. For these kind of perturbations from the set of equations (1-4)
one can obtain the following dispersion relation:

$$
{\omega}^2-k^2={{\omega_p^2}\over
{\gamma_0^3}}\left(1-{{k^2V_0^2}\over {{{\omega_c^2}/
{\gamma_0^6}}-\omega^2}}\right). \eqno (5)
$$

\begin{figure}{}
\begin{center}
\includegraphics[width=8cm]{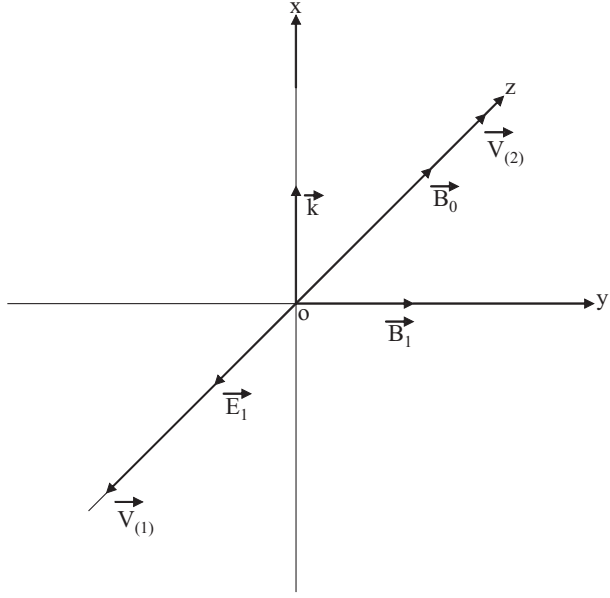}
\end{center}
\caption{Orientation of the perturbations under study.}
\end{figure}

Here $V_0$ and $\gamma_0$ are unperturbed velocity and
Lorentz-factor of particles, ${\omega}_c={{eB_{0}}/{m}}$ is the
cyclotron frequency ($B_0$ being the unperturbed magnetic field of
pulsar) and
${\omega}_p=\sqrt{4{\pi}\left({n_{0p}}+{n_{0b}}\right){e^2}/{m}}$
is the plasma frequency ($n_{0p}$ and $n_{0b}$ being the
unperturbed particle density of pulsar magnetosphere relativistic
electron-positron plasma and ultrarelativistic electron beam
respectively).

Equation (5) can be rewritten in the following form:

$$
a\omega^4+b\omega^2+c=0, \eqno (6)
$$
where

$$
a=\gamma_0^6, \eqno (7)
$$

$$
b=-\left(\omega_c^2+\omega_p^2\gamma_0^3+k^2\gamma_0^6\right),
\eqno (8)
$$

$$
c=k^2\omega_c^2+{{\omega_p^2\omega_c^2}\over
{\gamma_0^3}}-\omega_p^2\gamma_0^3k^2V_0^2. \eqno (9)
$$

From the expressions (6)-(9) one can conclude that if the
condition

$$
k^2V_0^2>{{k^2\omega_c^2}\over
{\omega_p^2\gamma_0^3}}+{{\omega_c^2}\over {\gamma_0^6}}, \eqno
(10)
$$
is fulfilled, than $\omega^2<0$ and $\omega$ is entirely
imaginary. The time dependence of perturbed quantities and namely
perturbed toroidal magnetic field is exponential $
B_1{\sim}exp\left(-i{\omega}t\right)$. Therefore, when the
condition (10) is fulfilled, then two-stream instability develops
in the magnetosphere of Crab Pulsar and exponentially growing
toroidal magnetic field is generated.

Substituting the parameters appropriate to Crab pulsar and its
magnetosphere one can easily find that the condition (10) is
really fulfilled.

As a result of the development of two-stream instabi\-lity
exponentially growing toroidal magnetic field is gene\-rated in
the magnetosphere of Crab pulsar (for other possible me\-chanisms
of toroidal magnetic field generation in the magnetosphere of Crab
Pulsar see [20] and [21]. The source of energy for the
gene\-ration of this field is the pulsar rotation slowing down.
Really, as it has been already mentioned above, as a result of
pulsar rotation together with frozen-in magnetic field, electric
field is generated. This electric field gets ener\-gy from pulsar
rotation slowing down. The electric field extracts electrons from
pulsar surface, accelerates them to ultrarelativistic velocities
and thus the ultrarelativistic electron beam is formed. In the
strong magnetic field of pulsar electron-positron pairs appear
from the beam particles and dense relativistic plasma of the
pulsar magnetosphere is formed. This plasma is penetrated by
ultrarelativistic electron beam. The system plasma-beam is
unstable and two-stream instability develops in it, as a result of
which toroidal magnetic field is generated in the pulsar
magnetosphere. The energy source for the generation of the
toroidal magnetic field is beam kinetic energy. The beam itself
acquires its kinetic energy from the electric field. As we have
been just mentioning the electric field is generated during pulsar
rotation together with frozen-in magnetic field and gets energy
from its rotation slowing down. Thus, the energy of the generated
toroidal magnetic field comes from pulsar rotation slowing down.

Superposition of generated toroidal magnetic field and pulsar
magnetic field will give the spiral configuration magnetic field.
Since plasma particles follow the magnetic field lines corotation
will be violated in the pulsar magnetosphere and instead of it we
will have differential rotation or shear flow of magnetospheric
plasma.

On large radial distances the step of the magnetic field spiral
should decrease and beyond the light cylinder magnetic field will
become practically purely toroidal. On larger radial distance from
pulsar - around $10^{17}$ cm this magnetic field is reconnected
with the magnetic field of Crab Nebula, which has also toroidal
structure (about one possible mechanism for the generation of this
field see [22]).

\vskip 15pt \noindent {\bf 4. Conclusions} \vskip 15pt Thus, in
the magnetosphere of Crab pulsar relativistic electron-positron
plasma is pene\-trated by ultrarelativistic electron beam. The
system plasma-beam is unstable and two-stream instability develops
in it. As a result exponentially growing toroidal magnetic field
is generated in the magnetosphere of Crab pulsar and after this
magnetic field structure changes to spiral. Since plasma particles
follow the magnetic field lines corotation will be violated. On
large radial distances step of the magnetic field spiral decreases
and we get practically purely toroidal magnetic field which is
finally reconnected with the magnetic field of Crab nebula.
Toroidal magnetic field is generated  in the pulsar magnetosphere
at the expense of energy released during pulsar rotation slowing
down.

\vskip 15pt \noindent {\bf References}

\begin{enumerate}

\item{F. Pacini, Nature, {\bf 216}, 467 (1967).}
\item{F. Pacini, Nature, {\bf 219}, 145 (1968).}
\item{T. Gold, Nature, {\bf 218}, 731 (1968).}
\item{T. Gold, Nature, {\bf 221}, 25 (1969).}
\item{P.A. Sturrock, Ap.J., {\bf 164}, 529 (1971).}
\item{D.B. Melrose, J. Astrophys. Astr., {\bf 16}, 137 (1995).}
\item{R.N. Manchester \& Taylor, J.H., "Pusalrs", W.H. Freeman
and company, San Francisco (1977).}
\item{F.G. Smith, "Pusalrs", Cambridge University Press,
Cambridge (1977).}
\item{P. Goldreich, \& W.H. Julian, Ap. J., {\bf 157}, 869 (1969).}
\item {J. Arons, in Proc. Workshop Plasma Astrophysics, pp.
273-286 (1981).}
\item{M.A. Ruderman, \& P.G. Sutherland, Ap.J., {\bf 196}, 51 (1975).}
\item{A.F. Cheng \& M.A. Ruderman, Ap.J., {\bf 235}, 576 (1980).}
\item{R. Buschauer \& G. Benford, M.N.R.A.S., {\bf 179}, 99 (1977).}
\item{G. Benford \& R.Buschauer, M.N.R.A.S., {\bf 179}, 189 (1977).}
\item{A.F. Cheng \& M.A. Ruderman, Ap.J., {\bf 212}, 800 (1977).}
\item{E. Asseo, R. Pellat \& M. Rosado, Ap.J., {\bf 239}, 661 (1980).}
\item{E. Asseo, R. Pellat \& M. Sol, Ap.J., {\bf 266}, 201 (1983).}
\item{V.N. Ursov \& V.V. Usov, Ap.S.S., {\bf 140}, 325 (1988).}
\item{V.V. Usov, Ap.J., {\bf 320}, 333 (1987).}
\item{T.A. Kahniashvili, G.Z. Machabeli \& I.S. Nanobashvili,, Phys. Plasmas, {\bf 4}, 1132 (1997).}
\item{I.S. Nanobashvili, Ap.S.S., {\bf 294}, 125 (2004).}
\item{G.Z. Machabeli, I.S. Nanobashvili \& M. Tendler, Physica Scripta, {\bf 60}, 601 (1999).}
\end{enumerate}

\end{document}